\documentclass[aps,prc,preprint,superscriptaddress,showpacs]{revtex4}

\usepackage{graphicx}

\begin{document}

\title{Direct neutron capture cross sections of $ ^{62} $Ni in the
$s$ process energy range}

\author{Thomas Rauscher}

\affiliation{Departement f\"ur Physik und Astronomie, Universit\"at Basel, 
CH-4056 Basel, Switzerland}

\email[]{Thomas.Rauscher@unibas.ch}
\homepage[]{http://nucastro.org}

\author{Klaus H. Guber}

\affiliation{Nuclear Science and Technology Division, Oak Ridge National
Laboratory, Oak Ridge, TN 37831, USA}

\altaffiliation{Current address: T\"UV Energie- und Systemtechnik GmbH
Baden-W\"urtemberg, D-68032 Mannheim, Germany}

\date{\today}

\begin{abstract}
Direct neutron capture on $^{62}$Ni is calculated in the DWBA and the
cross sections in the energy range relevant for s-process nucleosynthesis
are given. It is confirmed that the thermal value of the capture cross
section contains a subthreshold resonance contribution. Contrary to previous
investigations it is found that the capture at higher energies is dominated
by p-waves, thus leading to a considerably increased cross section at
s-process energies and a modified energy dependence.
\end{abstract}


\maketitle

The energy range of about 10 to 50 keV is relevant for neutron captures in
s-process nucleosynthesis in AGB stars and type II supernovae. A large number
of efforts have been undertaken to provide the necessary information on nuclear
cross sections and reaction rates for neutron captures on stable nuclides as
well as long-lived isotopes in s-process branchings. Nevertheless, in a
number of cases only thermal cross sections are known experimentally which have
been extrapolated to the energy region of interest. This is often done assuming
s-wave capture. However, there are two possible sources of error in such a treatment,
even in absence of resonances in the relevant energy region. Firstly, the direct
capture (DC) contribution may include higher partial waves which will 
become increasingly
important with increasing energy. Secondly, the thermal cross section may be
dominated by the tail of a subthreshold resonance. In the latter case, a simple
s-wave extrapolation is not justified. For neutron capture on $ ^{62} $Ni
both effects are relevant. A calculation of the DC contribution
can help to resolve both problems.

The two most recent recommended values of the 30 keV Maxwellian Averaged
Cross Section (MACS) are 35.5$ \pm 4 $
mb and 12.5$ \pm 4 $ mb, as quoted by Refs.\ \cite{bao87} and \cite{bao00},
respectively.
A closer scrutiny shows that they are both based
on the same experimental thermal neutron capture cross section but with 
different assumptions for the extrapolation to 30 keV \cite{beer01}. 
In the earlier extrapolation
the thermal value was taken to contain DC only which was then 
extrapolated
assuming s-wave capture. In the more recent re-evaluation it was realized that
a narrow subthreshold resonance at -0.077 keV could have contributed to the
thermal value of 14.5 barn. The new extrapolation was then performed by 
subtracting
an estimated resonance contribution and assuming s-wave DC for the
remaining cross section.

This clearly shows the need for a more consistent treatment, including a more
complete description of the DC process. In the following, the DC
contribution to the thermal cross section is studied as well as the
behavior in the range of several tens of keV.

The DC calculation was performed in a way as adopted
in many other investigations (e.g.\ \cite{rau,hari}), making use of
the code TEDCA \cite{krau}, including E1, M1, and E2 transitions. See
those references for more details on the DC formalism. The number
of open parameters is considerably reduced by employing folding potentials for
the optical potentials of target plus incident neutron and of the final neutron
bound state. The folding potential is obtained by folding the projectile and
target density distributions $ \rho  $$ _{a} $, $ \rho  $$ _{A} $
with an energy- and density-dependent effective nucleon-nucleon interaction
$ v_{\mathrm{eff}} $ \cite{kob}
\begin{equation}
V(R)=\lambda V_{F}(R)=\lambda \int \int \rho _{a}(\overrightarrow{r_{1}})\rho _{A}(\overrightarrow{r_{2}})v_{\mathrm{eff}}(E,\rho _{a},\rho _{A},s)\: d\overrightarrow{r_{1}}d\overrightarrow{r_{2}}\quad ,
\end{equation}
 with $ \lambda  $ being a potential strength parameter close to unity, and
$ s=\left| \overrightarrow{R}+\overrightarrow{r_{2}}-\overrightarrow{r_{1}}\right|  $,
where $ R $ is the separation of the centers of mass of the projectile and
the target nucleus. For nucleons the density distribution is usually
described by a delta function, simplifying the double integration to a
single one. 
Often, the imaginary part
of the potential is very small due to the small flux into other reaction 
channels
and can be neglected in most cases involving neutron capture at low energies
\cite{ohu}. In the present case this assumption is well justified because no
other reaction channels are open in the investigated energy range.
The density was derived from measured charge distributions \cite{vries}.

For the potential describing the final bound state of the captured neutron the
parameter $ \lambda _{\mathrm{f}} $ is fixed by the known binding energy.
The value of $ \lambda _{\mathrm{i}} $ in the entrance
channel is left to be determined. This can be achieved by describing 
elastic neutron scattering data.
In general, elastic scattering contains contributions from potential scattering
and resonant scattering. Since a description of the non-resonant direct process
is to be found, it is essential to use potential scattering cross sections or
scattering lengths without resonant contributions. 
Here, $ \lambda _{\mathrm{i}} $ is determined by reproducing
the potential scattering length of 6.3$ \pm 0.4 $ fm \cite{mug}. 
The experimental error can be included, leading to a range
of permitted values for $ \lambda  $$ _{\mathrm{i}} $ bounded by the two
extremes $ \lambda  $$ _{\mathrm{high}}=0.81552 $ and $ \lambda  $$ _{\mathrm{low}}=0.81392 $.
Thus, there remain no open parameters in this calculation which would have to
be fitted to reaction data.

Not only capture of neutrons into the ground state of $ ^{63} $Ni is relevant
but also transitions to excited states have to be included. Table 1 shows the
considered final states and transitions, along with the adopted spectroscopic
factors and the dominating partial waves. The level scheme of $ ^{63} $Ni
and the spectroscopic factors are taken from several $ ^{62} $Ni(d,p)$ ^{63} $Ni
experiments, as usual. The spectroscopic factors are renormalization factors
for each transition to a certain nuclear level, containing nuclear structure
effects. The error on the spectroscopic factors is given as about 20\%, therefore
an additional uncertainty of 20\% has been added to the error in $ \lambda  $$ _{\mathrm{i}} $
arising from the scattering lengths. The relative importance of the contribution
of each transition depends on the reaction $ Q $ value, the possible partial
waves (s, p, d, \ldots-wave capture) and the spectroscopic factors. Therefore,
nuclear states at high excitation energy do not contribute considerably, neither
do states with small spectroscopic factors. By far dominating are E1 s- and
p-wave captures. Higher partial waves and multipolarities have been included
in the calculation but as their contribution is lower by several ten orders 
of magnitude they are not explicitly considered in the discussion below.

\begin{table}
\caption{Transitions considered in the calculation of direct neutron
capture on $^{62}$Ni ($J^{\pi}=0^+$). Shown are the states in the final
nucleus $^{63}$Ni, the reaction $Q$ value, contributing angular momentum
$l$ and the spectroscopic factors $C^2S$.}
\begin{ruledtabular}
\begin{tabular}{rcrclr}
$Q$ [MeV] & $J^{\pi}$ & \multicolumn{3}{c}{Transition} & $C^2S$\\
\hline
6.838 & $\frac{1}{2}^-$ & s & $\longrightarrow$ & 2p$_{1/2}$ & 0.370\footnotemark[1] \\
      &                 & d & $\longrightarrow$ & 2p$_{1/2}$ & 0.370\footnotemark[1]\\
6.751 & $\frac{5}{2}^-$ & d & $\longrightarrow$ & 1f$_{5/2}$ & 0.563\footnotemark[1] \\
6.683 & $\frac{3}{2}^-$ & s & $\longrightarrow$ & 2p$_{3/2}$ & 0.275\footnotemark[1] \\
      &                 & d & $\longrightarrow$ & 2p$_{3/2}$ & 0.275\footnotemark[1] \\
6.323 & $\frac{3}{2}^-$ & s & $\longrightarrow$ & 2p$_{3/2}$ & 0.080\footnotemark[1] \\
      &                 & d & $\longrightarrow$ & 2p$_{3/2}$ & 0.080\footnotemark[1] \\
5.835 & $\frac{1}{2}^-$ & s & $\longrightarrow$ & 2p$_{1/2}$ & 0.330\footnotemark[1] \\
      &                 & d & $\longrightarrow$ & 2p$_{1/2}$ & 0.330\footnotemark[1] \\
5.511 & $\frac{3}{2}^-$ \footnotemark[2] & s & $\longrightarrow$ & 2p$_{3/2}$ & 0.125\footnotemark[1] \\
      &                 & d & $\longrightarrow$ & 2p$_{3/2}$ & 0.125\footnotemark[1] \\
3.885 & $\frac{1}{2}^+$ & p & $\longrightarrow$ & 3s$_{1/2}$ & 0.190\footnotemark[1] \\
3.665 & $\frac{5}{2}^-$ & d & $\longrightarrow$ & 1f$_{5/2}$ & 0.027\footnotemark[1] \\
3.555 & $\frac{5}{2}^+$ & p & $\longrightarrow$ & 2d$_{5/2}$ & 0.036\footnotemark[3] \\
3.129 & $\frac{5}{2}^+$ & p & $\longrightarrow$ & 2d$_{5/2}$ & 0.016\footnotemark[3] \\
2.926 & $\frac{5}{2}^+$ & p & $\longrightarrow$ & 2d$_{5/2}$ & 0.027\footnotemark[3] \\
2.899 & $\frac{5}{2}^+$ & p & $\longrightarrow$ & 2d$_{5/2}$ & 0.050\footnotemark[3] \\
2.786 & $\frac{5}{2}^+$ & p & $\longrightarrow$ & 2d$_{5/2}$ & 0.070\footnotemark[3] \\
2.580 & $\frac{1}{2}^+$ & p & $\longrightarrow$ & 3s$_{1/2}$ & 0.069\footnotemark[3] \\
2.462 & $\frac{5}{2}^+$ & p & $\longrightarrow$ & 2d$_{5/2}$ & 0.032\footnotemark[3] \\
2.294 & $\frac{5}{2}^+$ & p & $\longrightarrow$ & 2d$_{5/2}$ & 0.024\footnotemark[3]
\end{tabular}
\end{ruledtabular}
\footnotetext[1]{Excitation energies, spins, and
spectroscopic factors from \cite{anf}.}
\footnotetext[2]{Spin assignment from \cite{hut}.}
\footnotetext[3]{Excitation energies, spins, and spectroscopic factors
from \cite{staub}.}
\end{table}

The total DC cross section is the sum of all transitions to the
final states. Realizing that s-, p- and d-waves are dominant, a simple parameterization
of the cross section can be given:\begin{equation}
\label{eq:param}
\sigma =\frac{X}{\sqrt{E}}+Y\sqrt{E}+ZE^{3/2}\quad ,
\end{equation}
 with the three terms referring to transitions with orbital angular momentum
$ l= $0, 1, 2. Including the errors derived above, the following range $ \sigma _{\mathrm{low}}\leq \sigma \leq \sigma _{\mathrm{high}} $
is found:
\begin{eqnarray}
\label{eq:sighi}
\sigma _{\mathrm{high}} & =1.1\left( \frac{2.714\times 10^{-2}}{\sqrt{E}}
+3.85\times 10^{-3}\sqrt{E}
\right) \quad , & \\
\label{eq:siglow}
\sigma _{\mathrm{low}} & =0.9\left( \frac{2.48\times 10^{-2}}{\sqrt{E}}
+3.525\times 10^{-3}\sqrt{E}
\right) \quad ,
\end{eqnarray}
with the center-of-mass energy $ E $ of the incident neutron given in keV
and the resulting cross section $ \sigma  $ in barn. The factors before the
brackets describe the uncertainty in the spectroscopic factors. 
In the above Eqs.\ \ref{eq:sighi}-\ref{eq:siglow} the d-wave terms are
not shown because they are smaller by six orders of magnitude than the
p-wave terms.
With such a parameterization
it is easy to disentangle the contributions from different partial waves.
Contrary to
previous estimates, p-wave capture dominates the cross section in the energy
range relevant for s-process nucleosynthesis. Therefore, the cross section increases
with energy after an initial decline, leading to a higher value at 30 keV. This
contribution is entirely due to capture in the 1/2$ ^{+} $ state at $ E_{x}=2.955 $
MeV. This state is well established as it was seen in several
experiments \cite{anf,hut,staub}. A number of 5/2$ ^{+} $ states and 
a 1/2$ ^{+} $ state at higher
excitation energy are also reached by p-wave capture but their contributions
are small due to smaller $ Q $ values and smaller spectroscopic factors.

Since potential scattering lengths were used, the derived potential should not
include resonant contributions. In other words, the calculated thermal cross
section of 5.21$ \pm  $0.76 barn is the purely direct contribution to the
measured cross section of 14.5$ \pm  $4 barn whereas the difference is due
to the high-energy tail of the subthreshold resonance plus the
low-energy tails of resonances at positive energies. The DC value
also fits well into
the thermal cross section systematics of the other Ni isotopes which do
not show subthreshold resonances.

\begin{figure*}
\includegraphics[angle=-90,scale=0.5]{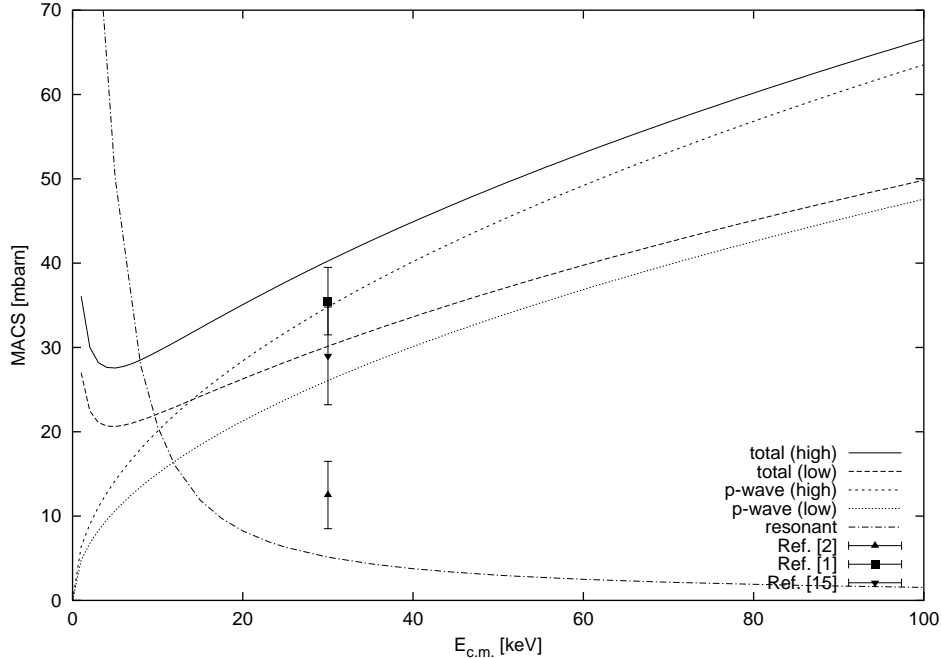}
\caption{\label{fig:cs}MACS
in the energy range $1\leq E_{\rm
c.m.}\leq 100$ keV. For the DC component, the upper and lower limit of the 
sum of all partial waves (total), and the high and low limit of the p-wave
are shown. The results are compared to two values derived from
thermal neutron capture data and extrapolated to 30 keV, taken from
\cite{bao87} and \cite{bao00}. 
A previous experimental value from \cite{bespe74}
is also shown.
Finally, the contribution of resonant capture including several
resonances and calculated from
resonance parameters taken from \cite{literature} is plotted (dot-dashed
line).}
\end{figure*}

It is quite simple to convert the above parameterization in Eq. (\ref{eq:param})
of the cross section $ \sigma  $ to one describing the MACS 
$ \left\langle \sigma \right\rangle = \left\langle \sigma v
\right\rangle / v_T $
by multiplying the p-wave term by 1.5, the d-wave term by 3.75, and keeping
the previous parameters, thus obtaining\begin{equation}
\left\langle \sigma \right\rangle _{\rm DC}=\frac{X}{\sqrt{E}}
+1.5Y\sqrt{E}+3.75ZE^{3/2}\quad .
\end{equation}
Thus, the DC MACS is considerably enhanced when including the p-wave capture.
This leads to a MACS of $ 30.1\leq \left\langle \sigma \right\rangle _{\rm
DC} \leq 40.2 $
mb at 30 keV. This is back to the original value given in 
\cite{bao87} (but for different reasons), also consistent with older data 
\cite{bespe74,bespe75}.
The resulting MACS up to 100 keV are shown in Fig.~\ref{fig:cs}. 

The resonant contribution can be estimated from the resonance parameters
given in literature \cite{literature} and is shown as
an additional line in Fig.\ 1. Although this may look close to a 1/v
function at first glance, the behavior is not that simple 
because it is a sum of several
resonances and the
energy-dependence of the widths has been taken into account in the
calculation performed with the multilevel $R$-matrix code SAMMY
\cite{larson}.
For a determination of the total rate,
it has to be added to the direct terms. As can be seen from Fig.\ 1,
the resonant contribution is small (5.13 mb) at 30 keV.
The fact that resonances are not very important at this energy is
underlined by a comparison to statistical model calculations
\cite{rau2000} which shows that the direct reaction mechanism is dominating.
The level density in this nucleus
is theoretically estimated to be close to the limit to allow
application of the statistical model.
Thus it is to be expected that any statistical model will even overestimate the
actual statistical resonant cross section. Single, previously unresolved
resonances could still contribute
but due to the rather wide level spacing their effects might be limited because
of the averaging taking place in the integration performed in the
calculation of the MACS and the reaction
rate. Nevertheless, a measurement in the relevant energy region is 
highly recommended.

Neglecting d-wave capture and utilizing the above parameterizations, the 
resulting astrophysical reaction rate for the direct contribution 
in cm$^3$ mol$^{-1}$ s$^{-1}$ is given by
\begin{equation}
N_{A}\left\langle \sigma v\right\rangle _{\rm DC} =\frac{2.64545\times 10^7}{\sqrt{\mu }}\left(X+1.2926\times 10^2\,Y\,T_{9}\right) \quad ,
\end{equation}
with the reduced mass $ \mu  $, Avogadro's number $ N_{A} $ and the temperature
$ T_{9} $ in 10$ ^{9} $ K. The constant term arises from s-wave
capture, the temperature dependent term includes the p-waves. The thermal
population of excited states in the target can be neglected at low temperatures.
The rate (also the cross section and the MACS) is bounded
by $ X_{\mathrm{low}}=2.232\times 10^{-2} $ barn keV$ ^{1/2} $, $ Y_{\mathrm{low}}=3.1725\times 10^{-3} $
barn keV$ ^{-1/2} $ and $ X_{\mathrm{high}}=2.985\times 10^{-2} $ barn
keV$ ^{1/2} $, $ Y_{\mathrm{high}}=4.235\times 10^{-3} $ barn keV$ ^{-1/2} $.
This rate is a lower limit because resonant contributions are neglected.
The rate obtained by fitting the resonant contribution shown in Fig.\ 1 has to
be added to the above DC rate:
\begin{eqnarray}
\label{eq:resrate}
N_{A}\left\langle \sigma v\right\rangle _{\rm res}&=&\exp \left(
-3.80993-\frac{0.0512461}{T_9}-\frac{1.63596}{T_9^{1/3}}+21.4001
T_9^{1/3}-3.29138 T_9 \right. \nonumber \\
&&\left. +0.3822399 T_9^{5/3}-5.52698\ln T_9 \right) \quad.
\end{eqnarray}
The fit function was chosen according to the standard format as
specified in \cite{rau2000}.
The resulting total rate $N_A \left\langle \sigma v\right\rangle =
N_{A}\left\langle \sigma v\right\rangle _{\rm DC}+
N_{A}\left\langle \sigma v\right\rangle _{\rm res}$ is valid in the
temperature range $0.001\leq kT\leq 100.0$ keV.

The enhanced rate has some impact also on type II supernova nucleosynthesis.
In a recent study of nucleosynthesis in type II supernovae, using an
extended reaction network including all stable nuclides up to Bi
\cite{rau01a,rau01b,rau01c}, the smaller value of \cite{bao00} was used.
Although AGB stars
are considered to be the main source of the s-process nuclides in the solar
system (see e.g.\ \cite{bu}), considerable 
s-processing also occurs in the early burning stages of
massive stars producing the weak s-process component. 
(This component also provides seed nuclides for the $ \gamma  $-process
which produces proton-rich nuclei by photo-disintegration.)
In the supernova nucleosynthesis calculation an enhanced
production of $ ^{62} $Ni relative to the other Ni isotopes was found (see
Fig.\ 2 in \cite{rau01b} and Figs.\ 2--6 in \cite{rau01c}). 
Using the previous value for $ ^{62} $Ni(n,$ \gamma  $)$ ^{63} $Ni
as given by \cite{bao87} -- being higher by about a factor of 3 --
there is no such overproduction (see e.g.\ \cite{hoffm01}).
Using the rate proposed here in type II supernova nucleosynthesis calculations,
a similar result can be expected because the new rate is -- within
errors -- compatible with the previous value of \cite{bao87}.

Summarizing, the consistent treatment of 
the reaction $ ^{62} $Ni(n,$ \gamma  $)$ ^{63} $Ni
in the DC formalism confirmed the importance of a subthreshold 
contribution
to the thermal cross section. However, it also showed that p-wave capture is
the dominating contribution to the cross sections and rates in the energy range
relevant for s-process nucleosynthesis, contrary to earlier assumptions. The
new rate leads to $ ^{62} $Ni abundances in type II supernovae consistent
with the abundance levels of neighboring isotopes and curing a
pronounced overproduction observed in some of the simulations. 
This underlines the fact
that thermal neutron capture cross sections often cannot be simply extrapolated 
to the 30 keV region by assuming s-wave capture. On the other hand, the example
of $ ^{62} $Ni also shows that even thermal cross sections should be evaluated
including a DC analyses to disentangle resonance effects and direct
contributions.

The remaining uncertainty in the cross sections can be reduced by resonance
measurements in the relevant energy region to include possible effects of single
resonances, by neutron scattering experiments with improved accuracy to reduce
the error in the potential scattering lengths, and by activation experiments
within the relevant energy region which include direct and resonant 
contributions
simultaneously but can currently only probe the MACS at a limited number of
energies. In compilations such as \cite{bao00} it would also be advantageous
to identify those rates which are based only on the capture of thermal neutrons
or are dominated by extrapolated terms.

This work is partially supported by the Office of Environmental
Management and the Office of Science, U.S. Department of Energy
(contract no.\ DE-AC05-00OR22725 with UT-Battelle, LLC) and
by the Swiss National Science Foundation (grant 2000-061822.00).
T.R. acknowledges a PROFIL professorship of the Swiss NSF (grants 
2124-055832.98, 2124-067428.01).

\end{document}